\documentclass[final,5p,times]{elsarticle}
\pdfoutput=1 
\usepackage{graphicx}
\usepackage{float}
\usepackage{tabularx}
\usepackage{adjustbox}
\usepackage{lipsum}
\usepackage{bbm}
\usepackage[dvipsnames]{xcolor,colortbl}
\usepackage[labelfont=bf,labelsep=period]{caption}
\usepackage{booktabs}
\usepackage{amsmath,mathrsfs,lineno,amssymb,,amsbsy}
\usepackage{natbib}

\biboptions{sort&compress}

\journal{arxiv}

\usepackage{etoolbox}
\patchcmd{\keyword}{\textit}{}{}{}

\begin{document}

\begin{frontmatter}

\title{\textbf{Piezoelectric metastructures for simultaneous broadband energy harvesting and vibration suppression of traveling waves}}

\author[]{\textbf{Z. Lin}}
\author[]{\textbf{H. Al Ba'ba'a}}
\author[]{\textbf{S. Tol}\corref{mycorrespondingauthor}}
\cortext[mycorrespondingauthor]{Authors to whom any correspondence should be addressed.}

\address {\normalfont{Department of Mechanical Engineering, University of Michigan, Ann-Arbor, MI 48109-2125, United States}}
\address {\normalfont{E-mail:stol@umich.edu}}
\begin{abstract}
In this paper, we explore an electromechanical metastructure consisting of a periodic array of piezoelectric bimorphs with resistive-inductive loads for simultaneous harvesting and attenuation of traveling wave energy. We develop fully coupled analytical models, i.e., an electroelastic transfer matrix method, and exploit both locally-resonant and Bragg band gaps to achieve a multifunctional metastructure which is capable for maximum energy conversion and vibration mitigation in a broadband fashion. Our analytical and numerical results show that the proposed metastructure can achieve energy harvesting efficiency up to 95$\%$ at the local resonance frequency of $3.18$ kHz, while reaching about 51$\%$ at $5.8$ kHz near the upper limit of the Bragg band gap. The broadband vibration mitigation performance based on 50$\%$ power attenuation is predicted as $1.8$ kHz and $1.1$ kHz in the vicinity of the band gaps. The theoretical frameworks and the applicability of the proposed metastructure are validated using a full-scale experimental setup. 

\end{abstract}

\end{frontmatter}

\section{Introduction}
\sloppy
Piezoelectric energy harvesting has attracted an increasing research interest in the past few decades due to their potential for powering small electronics, ease of fabrication at different geometric scales, and high power density compared to other common transduction mechanisms  \cite{Anton2007A2003-2006,Cook-Chennault2008PoweringSystems,Saadon2011AHarvesters,Liu2018AApplications,Safaei2019A20082018,Covaci2020PiezoelectricReview}. The harvesting of standing waves and vibrations has been extensively researched through the use of linear and nonlinear energy harvesters, such as bimorph cantilevers with piezoelectric layers undergoing base excitation \cite{Erturk2009AnExcitations,erturk2011broadband,ferrari2009improved,mei2020tri,zhou2018numerical,yeo2016efficient}. Researchers also studied flow energy harvesting through aeroelastic \cite{bibo2012energy,dunnmon2011power} and hydroelastic \cite{akcabay2012hydroelastic} vibrations. While the existing efforts mostly investigate the harmonic and/or random excitation and focus on the resonant phenomenon to achieve peak energy harvesting capabilities, the potential of piezoelectric harvesters has not fully explored for harvesting energy from traveling elastic waves. Tol et al. \cite{Tol2016PiezoelectricEnhancement} developed a fully coupled electromechanical model of a bimorph piezoelectric patch harvester and proposed a wave propagation approach enabling maximum energy conversion in traveling elastic wave scenarios including transient excitations. In this paper, our primary goal is to extend the wave-based method to periodic piezoelectric harvester settings and exploit the extraordinary properties of metamaterials and phononic crystals (PC) to achieve enhanced harvesting performance with maximum efficiency in a broadband fashion.     

Metamaterials and PCs are periodic structures consisting of the locally resonating unit structures and spatially modulated elastic moduli/mass density lattices, respectively, and they can manipulate elastic wave propagation in non-traditional ways \cite{Hussein2014,jin2019gradient,srivastava2015elastic}. For instance, they can forbid the propagation of elastic/acoustic waves in the certain frequency bands (a.k.a. band gaps). Band gap mechanism in metamaterials is based on the locally resonating structures while Bragg band gap is induced by the constructive and destructive interferences as a result of the spatial periodicity of PCs. The ability to tune the performance of metamaterials and phononic crystals without requiring structural modifications is one of the challenges in the wave control field. In an effort to overcome this limitation, researchers embedded periodic arrays of piezoelectric transducers in the elastic waveguide and tune its dispersion behavior via external electrical load on the piezoelectric array. Casedei et al. \cite{casadei2012piezoelectric} showed the evidence of the tunable resonant-type band gap resulting from the resonance characteristics of the shunts leading to strong attenuation and to negative group velocities at frequencies defined by the circuits’ inductance. Later, piezoelectricity is studied for reconfiguring the wave directivity in active metamaterials \cite{Celli2015}, tuning the gradient-index lenses and metasurfaces \cite{yi2016flexural}, controlling band gap mechanism over a wide frequency range for vibration absorption \cite{chen2014piezo,SUGINO2018323}, wave filtering \cite{Trainiti2019Time-PeriodicExperiment}, and for spatiotemporally modulated non-reciprocal structures \cite{Marconi2020ExperimentalArray,Sugino2020NonreciprocalStrategies}. In these studies, inverse piezoelectric effect was exploited in periodic arrays to modify their effective modulus and control wave propagation. 

On the other hand, researchers also exploited the periodic structures in order to enhance piezoelectric energy harvesting based on the direct piezoelectric effect. For instance, efficiency of electrical power generation is further improved by developing proper strategies for spatial localization of wave energy at the harvester location via defect mode PCs \cite{Gonella2009,Wang_2010}, elastic mirrors \cite{tol2017structurally,carrara2013metamaterial}, gradient-index lenses \cite{Tol2016Gradient-indexHarvesting,Tol2017PhononicHarvesting,Tol20193D-printedHarvesting}, and metasurfaces with focusing \cite{lin2021elastic} and rainbow trapping \cite{DePonti2020GradedHarvesting,Chaplain2020DelineatingHarvesting} effects. Moreover, researchers studied piezoelectric metastructures for simultaneous vibration suppression and energy harvesting of standing waves in cantilever type structures with harmonic base excitation \cite{Hu2018InternallyHarvesting,Sugino2018AnalysisHarvesting,Chen2019AHarvesting,Alshaqaq2020GradedHarvesting}. For instance, in an effort for broadband harvesting of the low frequency vibration energy, Chen et al. \cite{CHEN20135} explored piezoelectric patch array with heavy masses based on the PC concept. Later, Sugino and Erturk \cite{Sugino2018AnalysisHarvesting} studied local resonators in the form of piezoelectric cantilever harvesters with tip mass attachment for energy conversion of standing waves. Both approaches are promising for low frequency broadband harvesting with the cost of adding a significant mass to the structure. More recently, Alshaqaq and Erturk \cite{Alshaqaq2020GradedHarvesting} proposed a locally-resonant metastructure with a graded array of shunted piezoelectric patches on a cantilever beam analyzed by a modal analysis framework. Unlike previous studies, in this paper, we introduce an electromechanical metastructure (EMM) for harvesting energy from travelling elastic waves and exploit the concepts of locally-resonant metamaterials and phononic crystals within the same structure. Specifically, we develop a new wave-based theoretical framework (i.e., fully coupled electroelastic transfer matrix method) combined with a power flow approach to enable the maximum harvesting and vibration attenuation performance. We analyze the wideband multifunctional EMM performance by coupling the locally-resonant and Bragg band gaps. Furthermore, our analyses offer a design space for geometric scales of the piezoelectric metastructure while maintaining peak conversion efficiency at different frequencies without any parasitic masses. Theoretical models and the proposed EMM are successfully validated through numerical simulations and experiments. 

The outline of this paper is as follows: In section~\ref{analytical_model}, the analytical model is developed based on transfer matrix method (TMM) to study the fully coupled EMM system. Based on this model, we investigate the energy harvesting and wave attenuation performance of the EMM in section~\ref{sec:performance}. Furthermore, the multi-functional EMM is experimentally validated in section~\ref{sec:experiment_validation}. Section~\ref{sec:concluding} summarizes the key points in this study.

\section{Wave characteristics of an infinite electromechanical waveguide}
\label{analytical_model}
\subsection{Electromechanical model}

\begin{figure*}[]
     \centering
\includegraphics[]{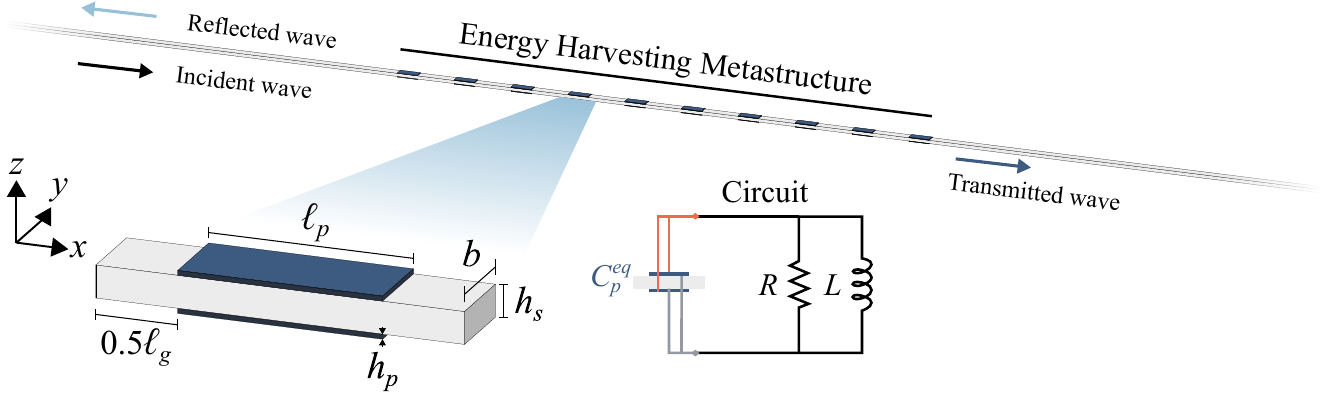}
     \caption{Infinite aluminum beam waveguide with periodically attached shunted piezoelectric patches (marked as the energy harvesting metastructure). Each electromechanical unit cell of the metastructure is connected to a resistive-inductive loading in a parallel connection.}
     \label{fig:infinite_beam}
\end{figure*}

We consider an infinitely long beam with a width of $b$ and thickness of $h_s$ as the host waveguide (i.e., substrate). Of particular interest in this study is the flexural wave propagation, which is captured by the transverse displacement, $w(x,t)$, of the beam. Denoting $\lambda$ as the wavelength of the propagating flexural wave and assuming a relatively small thickness (i.e., $\lambda/h_s>10$), the equation of motion is governed by Euler-Bernoulli theory:
\begin{equation}
    YI \frac{\partial^4 w}{\partial x^4} + m \frac{\partial^2 w}{\partial t^2} = 0
\end{equation}
where $YI$ represents the bending stiffness and $m$ is the mass per unit length. At the mid span of the beam, an array of bimorph piezoelectric patches are periodically placed with a spacing of $\ell_g$. Each patch has a length of $\ell_p$, width of $b$, and thickness of $h_p$ (see figure \ref{fig:infinite_beam}). The piezoelectric patches are connected to a resistive-inductive circuit in parallel connection. As a result, an electromechanical resonance occurs in the vicinity of the electrical circuit resonance:
\begin{equation}
    \omega_t^2=\frac{1}{L C_{p}^{eq}}; \hspace{0.5cm} C_{p}^{eq}=\frac{2\varepsilon_{33}^{s}b \ell_{p}}{h_{p}}
    \label{eq:inductance_resonance}
\end{equation}
where $\omega_t$ is the targeted resonant frequency of energy harvesting, $L$ is the inductive loading, $\varepsilon_{33}^{s}$ is the piezo electric permittivity, and $C_{p}^{eq}$ is the equivalent capacitance of the piezoelectric bimorph with parallel connection. Derivation of the fully coupled electromechanical model with the presence of piezoelectric bimorphs can be found in literature (for example, references \cite{Erturk2009AnExcitations,Tol2016PiezoelectricEnhancement,erturk2011piezoelectric}) and are not repeated here for the sake of brevity. Instead, we present the coupled moment equilibrium equation at the right and left boundaries of the $i^{\text{th}}$ bimorph (denoted with superscripts R and L, respectively):
\begin{subequations}
\begin{equation}
    M_{i}^{\text{L}} = M_{i-1}^{\text{R}} + \chi v_i
    \label{eq:moment_eq_1}
\end{equation}
\begin{equation}
    M_{i+1}^{\text{L}} = M_i^{\text{R}} - \chi v_i
    \label{eq:moment_eq_2}
\end{equation}
\label{eq:force_eq_1}
\end{subequations}

\sloppy 
\noindent
where $\chi=be_{31}(h_{p}+h_{s})$ is the backward coupling term, and $e_{31}$ is the effective piezoelectric stress constant. The voltage across the $i^\text{th}$ piezoelectric bimorph, $v_i$, is defined as \cite{Tol2016PiezoelectricEnhancement}:
\begin{equation}
   v_i = \gamma \left(\partial_x w_i^\text{L} - \partial_x w_i^\text{R} \right) 
   \label{eq:voltage_disp}
\end{equation}
\begin{subequations}
\begin{equation}
    \gamma = \frac{\textbf{i} \omega z \chi}{\textbf{i} \omega C_{p}^{eq} z + 1}
\end{equation}
\begin{equation}
    z = \left(\frac{1}{R} + \frac{1}{\mathbf{i}\omega L} \right)^{-1}
    \label{eq:impedence_freq}
\end{equation}
\end{subequations}
where $z$ is the electrical impedance of the resistive-inductive loading in parallel configuration. 

\subsection{Transfer matrix for the electromechanical unit cell}
\label{sec:Transfer_Matrix_for_Electromechanical_Unit_Cell}
Owing to the periodic nature of metastructures, it is customary to conduct unit cell analysis to reveal the wave propagation patterns (i.e., dispersive characteristics), including frequency band gaps. As the voltage and impedance are more conveniently expressed in frequency domain, a suitable methodology for such electromechanically coupled analysis is TMM, which relates state vectors between two points on a structure at a given excitation frequency $\omega$. For flexural waves, the state vector is constituted of the transverse displacement ($w$), angular displacement ($\phi$), moment ($M$), and shear force ($F$). Dispersion relation is obtained by mapping the eigenvalues $\zeta$ of the metastructure transfer matrix to the complex wavenumbers \cite{Hussein2014}. While TMM have been previously developed based on modeling piezoelectric/substrate composite by expressing its modulus as a complex value \cite{airoldi2011design}, we approach the problem differently and develop a point moment transfer matrix to account for the voltage generated in the electromechanical unit cell, due to the fact that the electromechanical coupling effect of the piezoelectric composite introduces additional moment terms at its boundaries.  

\begin{figure}[]
     \centering
     \includegraphics[]{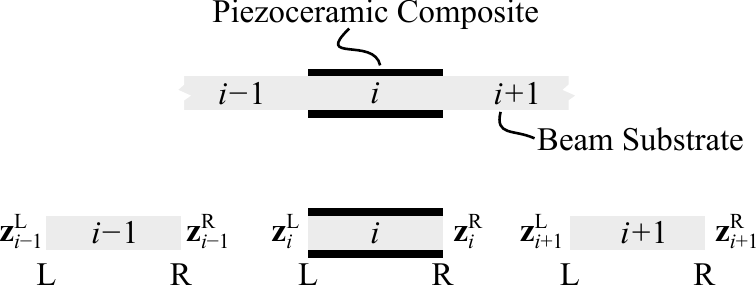}
     \caption{Schematics of the state vector definitions in the transfer matrix method.}
     \label{fig:Sch_TMM}
\end{figure}

For the $i^{\text{th}}$ section of the beam with an arbitrary length $x$, the transfer matrix relating the state vectors at right $\mathbf{z}^\text{R}_{i}$ and left $\mathbf{z}^\text{L}_{i}$ boundaries of that section is given by (see figure~\ref{fig:Sch_TMM}):
\begin{equation}
\mathbf{z}^\text{R}_{i} = \mathbf{T} \mathbf{z}^\text{L}_{i}; \hspace{0.5cm} \mathbf{z}_{i} = 
\left\{{w}_i, {\phi}_i, {M}_i, {F}_i\right\}^T
    \label{eq:TM_beam_Gen}
\end{equation}
where the components of the transfer matrix are given by:
\begin{equation}
    \mathbf{T}(x) =
    \begin{bmatrix}
    \mathbf{T}_{11}(x) & \mathbf{T}_{12}(x)\\ \mathbf{T}_{21}(x) & \mathbf{T}_{22}(x)
    \end{bmatrix}=
    \mathbf{E}
    \mathbf{Q} \mathbf{U} \mathbf{\Lambda}(k,x) \mathbf{U}^{-1} \mathbf{Q}^{-1} \mathbf{E}^{-1}
\end{equation}
with
\begin{subequations}
\begin{equation}
    \mathbf{U} = \frac{1}{2}
    \begin{bmatrix}
    1 & 1 & 1 & 1 \\ 
    -\textbf{i} & -1 & \textbf{i} & 1 \\
    -1 & 1 & -1 & 1 \\ 
    \textbf{i} & -1 & -\textbf{i} & 1 \\
    \end{bmatrix}
\end{equation}
\begin{equation}
    \mathbf{\Lambda}(k,x) = \mathbf{diag} [e^{-\textbf{i} kx} e^{-kx} e^{\textbf{i} kx} e^{k x}]
\end{equation}
\begin{equation}
    \mathbf{E} = 
    \mathbf{diag}
    \left[
    \begin{matrix} 
    1 & 1 & YI & YI
    \end{matrix}
     \right]
\end{equation}
\begin{equation}
    \mathbf{Q} = \mathbf{diag}
    \left[
    \begin{matrix} 
    1 & k & k^2 & k^3
    \end{matrix}
     \right]
\end{equation}
\end{subequations}
where $\textbf{i}$ is the imaginary unit and $k$ is the wavenumber defined as:
\begin{equation}
    k = \sqrt[4]{\frac{m \omega^2}{YI}}
\end{equation}

The bending stiffness, $YI$, and mass per unit length, $m$, of the beam substrate and piezoelectric composite sections are given by:
\begin{align}
\label{eqn:26}
YI=\left\{
\begin{aligned}
&\frac{Y_{s}bh_{b}^{3}}{12}  & & \text{Substrate} \\
&\frac{2b}{3}\left(Y_{s}\frac{h_{s}^{3}}{8}+c_{11}^{E}\left[\left(h_{p}+\frac{h_{s}}{2}\right)^{3}-\frac{h_{s}^{3}}{8} \right]\right)  & & \text{Composite}
\end{aligned}
\right.
\end{align}
\begin{align}
\label{eqn:25}
m=\left\{
\begin{aligned}
&\rho_{s}bh_{s} & & \text{Substrate} \\
&\rho_{s}bh_{s}+2\rho_{p}bh_{p} & & \text{Composite}
\end{aligned}
\right.
\end{align}
where $Y_s$ is the Young's modulus of substrate, $c_{11}^{E}$ is the elastic modulus of piezoceramic at constant electric field, $\rho_{s}$ and $\rho_{p}$ are the densities of the substrate and piezoelectric patch, respectively. 

Note that, the state vector of the left boundary of a waveguide's section can be found from the right boundary state vector via the inverse of the transfer matrix, which is evaluated as:
\begin{equation}
    \mathbf{T}^{-1}(x) = \mathbf{T}(-x)
    \label{equ:inverse_T}
\end{equation}

For the rest of the derivations, we distinguish between the transfer matrix of the piezoelectric composite and the substrate sections by the subscript $p$ and $s$, such that their transfer matrices are $\mathbf{T}_s$ and $\mathbf{T}_p$.

To relate the input and output waves passing through the ${i}^{\text{th}}$ piezoelectric composite section, we derive two transfer matrices for the additional moment effect caused by the electromechanical coupling. Using the relationship of the voltage to the angular displacements of the piezoelectric boundaries in equation (\ref{eq:voltage_disp}), the moment equilibrium equations can be expressed as:
\begin{subequations}
\begin{equation}
    M_i^{\text{L}} = M_{i-1}^{\text{R}} + \chi \gamma (\phi_i^\text{L}- \phi_i^\text{R})
    \label{eq:moment_eq_11}
\end{equation}
\begin{equation}
    M_{i+1}^{\text{L}} = M_{i}^{\text{R}} - \chi\gamma (\phi_i^\text{L} - \phi_i^\text{R})
    \label{eq:moment_eq_22}
\end{equation}
\label{eq:force_eq_11}
\end{subequations}

Subsequently,~linear and angular displacement compatibility and moment and force equilibrium conditions at the left-hand substrate/piezoelectric composite interface of the $i^{\text{th}}$ composite section can be written in the following matrix form:
\begin{equation}
    \begin{bmatrix}
    \mathbf{I} & \mathbf{0} \\ 
    -\mathbf{S} & \mathbf{I}\\
    \end{bmatrix}
    \mathbf{z}^\text{L}_{i} 
    =
    \mathbf{z}^\text{R}_{i-1} 
    -
    \begin{bmatrix}
    \mathbf{0} & \mathbf{0} \\ 
    \mathbf{S} & \mathbf{0}\\
    \end{bmatrix}
    \mathbf{z}^\text{R}_{i} 
\end{equation}
where
\begin{equation}
    \mathbf{S} = 
    \begin{bmatrix}
    0 & \chi \gamma \\
   0 & 0\\
    \end{bmatrix}
\end{equation}

Making use of the transfer matrix $\mathbf{T}_p$ (with $x = \ell_p$), we obtain the point moment transfer matrix ${\mathbf{P}_1}$ for the first interface:
\begin{equation}
\mathbf{z}^\text{L}_{i} =
    \underbrace{
    \begin{bmatrix}
    \mathbf{I} & \mathbf{0} \\ 
    \mathbf{S} \mathbf{T}_{p,11}(\ell_p) -\mathbf{S} & \mathbf{S} \mathbf{T}_{p,12}(\ell_p) + \mathbf{I}\\
    \end{bmatrix}^{-1}}_{\mathbf{P}_1}
    \mathbf{z}^\text{R}_{i-1} 
\end{equation}

Next, we obtain the point moment transfer matrix for the right-hand substrate/piezoelectric composite interface of the $i^{\text{th}}$ composite section, where the compatibility and equilibrium conditions can be expressed as:
\begin{equation}
\mathbf{z}^\text{L}_{i+1} = 
    \begin{bmatrix}
    \mathbf{I} & \mathbf{0} \\ \mathbf{S} & \mathbf{I}
    \end{bmatrix}\ \mathbf{z}^\text{R}_{i}
    -
    \begin{bmatrix}
    \mathbf{0} & \mathbf{0} \\ \mathbf{S} & \mathbf{0}
    \end{bmatrix}\
    \mathbf{z}^\text{L}_{i}
\end{equation}

Then, using equation~(\ref{equ:inverse_T}) we can derive the point moment transfer matrix $\mathbf{P}_2$ for the second interface as:
\begin{equation}
\mathbf{z}^\text{L}_{i+1} =
    \underbrace{
    \begin{bmatrix}
    \mathbf{I} & \mathbf{0} \\ \mathbf{S} - \mathbf{S} \mathbf{T}_{p,11}(-\ell_p) & \mathbf{I} - \mathbf{S} \mathbf{T}_{p,12}(-\ell_p)
    \end{bmatrix}}_{\mathbf{P_2}}
    \mathbf{z}^\text{R}_{i}
\end{equation}

Hence, the transfer matrix for the $i^{\text{th}}$ piezoelectric composite section, relating the left boundary of the $({i+1})^{\text{th}}$ section to the right boundary of the $({i-1})^{\text{th}}$ section becomes:

\begin{equation}
    \mathbf{z}^\text{L}_{i+1}
    = \mathbf{P}_2 \mathbf{T}_p(\ell_p) \mathbf{P}_1
    \mathbf{z}^\text{R}_{i-1}
    \label{eq:T2_z1_z3}
\end{equation}

Finally, the transfer matrix of the electromechanical unit cell shown in figure~\ref{fig:infinite_beam} can be conveniently expressed as:
\begin{equation}
    \mathbf{T}_{
    \text{EM}} = \mathbf{T}_s(\ell_g/2) \mathbf{P}_2 \mathbf{T}_p(\ell_p) \mathbf{P}_1 \mathbf{T}_s(\ell_g/2)
    \label{eq:TMM_UC}
\end{equation}

\subsection{Dispersion relations}

\begin{figure*}[]
     \centering
     \includegraphics[]{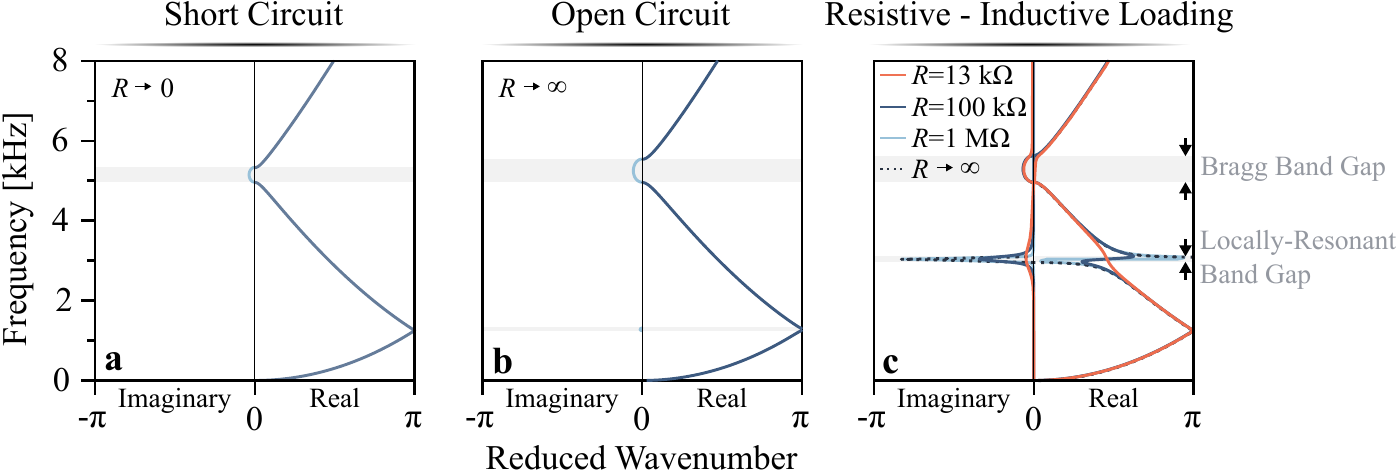}
     \caption{Dispersion relations for the electromechanical unit cell at different electrical loading conditions: (\textbf{a}) short and (\textbf{b}) open circuit, (\textbf{c}) resistive-inductive loading with various values of resistance showing both locally-resonant and Bragg band gaps.}
     \label{fig:Dispersion_Relations}
\end{figure*}

Dispersion relations carry essential information on the wave propagation characteristics by expressing the wavenumber as a function of frequency.~Here, we evaluate the dispersion relation for the unit cell by computing the eigenvalues of the transfer matrix given in equation~(\ref{eq:TMM_UC}), which are then mapped to the wavenumber. For the geometrical, mechanical, and piezoelectric properties presented in table~\ref{table_data}, figure~\ref{fig:Dispersion_Relations} shows the dispersion relations for different circuit configurations and conditions. As a first step, we investigate the metastructure at the short and open circuit conditions, i.e. with an electrical impedance of $z = 0$ and $z=100$ $\text{M}\Omega \approx \infty$, respectively. In both cases, narrow Bragg band gaps open due to the periodic variation in the mechanical properties. On the other hand, for pure inductive loading, the impedance reduces to $z = \textbf{i}\omega L$ 
resulting in a locally-resonant band gap centered around the electrical resonance of the circuit ($\omega_t=(LC^{eq}_p)^{-\frac{1}{2}}$).  When a resistor is connected in parallel with the inductor, the locally-resonant band gap closes due to the dissipative effects. It is important to note that the strong attenuation parameter (i.e., the imaginary wavenumber) is a consequence of the localization of energy in the electromechanical resonators, which is exploited for enhanced conversion efficiency as discussed in detail in section~\ref{sec:single_patch_opt}.

\begin{table}[]
\caption{Parameters of the EMM (Piezoelectric properties are for PZT-5H type).}
\label{table_data}
\centering
\begin{tabular}{l l l l}
\toprule
Parameter  & Value & Parameter & Value  \\
\midrule
$h_s$ & 2 mm &  $h_p$  & 0.3 mm\\
$b_s$ & 5 mm &  $b_p$ & 5 mm\\
$\ell_g$ & 35.5 mm & $\ell_p$ & 25 mm\\
$\rho_s$ & 2700 kg/$\text{m}^3$ & $\rho_p$ & 7500 kg/$\text{m}^3$\\
$Y_s$ & 70 GPa & $c_{11}^E$ & 60.6 GPa\\
$C_{p}^{eq}$ (Measured)& 12.5 nF & $e_{31}$ & -16.6 C/m\\
\bottomrule
\end{tabular}
\end{table}

\subsection{Fully coupled transfer matrix of the metastructure}
\label{Sec:TMM_Finite_MM}
We extend the transfer matrix formulation of an individual unit cell to analyze the elastrodynamics of the EMM in figure~\ref{fig:infinite_beam} as part of an infinite waveguide. Recall that the piezoelectric patches are equally spaced on the substrate with a gap distance of $\ell_g$, which has a transfer matrix denoted as $\mathbf{T}_s(\ell_g)$ (or $\mathbf{T}_s$ for brevity). Therefore, and for $n$ number of patches, the total transfer matrix $\mathbf{T}_n$ can be found via a successive multiplication of the transfer matrix corresponding to the $2n-1$ different sections ($n$ piezoelectric composit and $n-1$ beam substrate sections):
\begin{equation}
    \mathbf{T}_n = \mathbf{\hat{T}}_p (\mathbf{T}_s \mathbf{\hat{T}}_p )^{n-1}; \hspace{0.5 cm} \mathbf{\hat{T}}_p = \mathbf{P}_2 \mathbf{T}_p(\ell_p) \mathbf{P}_1
\end{equation}
which relates the state vectors in the incident and transmitted regions of the wave guide:
\begin{equation}
   \mathbf{z}^\text{L}_{2n}
    = \mathbf{T}_n \mathbf{z}^\text{R}_{0}
    \label{eq:Tn}
\end{equation}

Of specific interest here is to calculate the state vectors at the boundaries of the piezoelectric composite sections to compute the voltage output and perform power flow analysis. For instance, the left-hand and right-hand state vectors of the $(2i-1)^{\text{th}}$ piezoelectric bimorph can be found as a function of the input wave profile such that:
\begin{subequations}
\begin{equation}
   \mathbf{z}^\text{L}_{2i-1}
    = \mathbf{P}_1 (\mathbf{T}_s \mathbf{\hat{T}}_p)^{i-1} \mathbf{z}^\text{R}_{0}
\end{equation}

\begin{equation}
    \mathbf{z}^\text{R}_{2i-1}
    = \mathbf{T}_p \mathbf{z}^\text{L}_{2i-1}; \hspace{0.5cm} i = 1,2, \dots, n
\end{equation}
\end{subequations}
\subsection{Transmission and reflection coefficients}
The complete solution of the Euler-Bernoulli beam equation has two propagating components and two evanescent components, which can be expressed as:
\begin{equation}
    w_r = a_{1,r} e^{-\textbf{i} k_r x} + a_{2,r} e^{-k_r x} + a_{3,r} e^{\textbf{i} k_r x} + a_{4,r} e^{k_r x} 
    \label{eq:general_sol_disp}
\end{equation}
where $a_{1,r}, a_{2,r}, a_{3,r}, \text{and}$ $a_{4,r}$ are the complex wave amplitudes and $r = 0,1,\dots,2n$ denote the total number of the composite and substrate sections. The negative (positive) sign of the exponents denotes the forward (backward) propagating and evanescent waves. The complex wave amplitudes can be simultaneously solved for each section of the EMM by properly setting up the compatibility and equilibrium equations at each interface. Hence, one can obtain the transmission and reflection coefficients in different metastructure sections. Alternatively, we exploit TMM to calculate the transmission and reflection coefficients of the propagating waves. The use of TMM provides a convenient means for establishing unit cell analysis as well as analyzing a structure with a finite array of unit cells. In addition, it can be used in any structural configuration, whether periodic or aperiodic, with high computation accuracy. Another advantage is that TMM eliminates the need for formulating the linear and angular displacement continuity and moment and force equilibrium equations at the different interfaces in wave-based analyses.

\begin{figure*}[]
     \centering
     \includegraphics[]{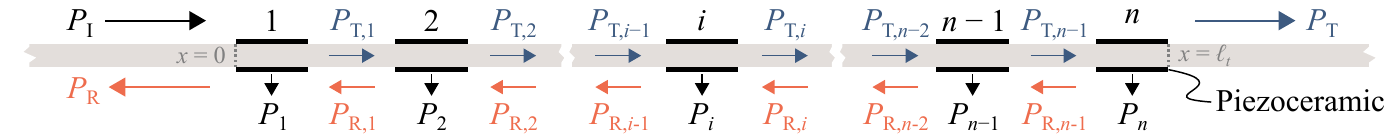}
     \caption{Illustration of the power flow in the piezoelectric metastructure, showing its different components, namely the incident ($P_{\text{I}}$), reflected ($P_{\text{R}}$), transmitted ($P_{\text{T}}$), and harvested ($P_{i}$) powers.}
     \label{fig:each_patch_power}
\end{figure*}

It is well established that Euler-Bernoulli theory relies only on the transverse displacement to compute the angular displacement, moment and shear forces. For instance, the wave at the input segment of the beam can be expressed as:

\begin{equation}
   \mathbf{z}^\text{R}_{0}
    =2\mathbf{E}_s \mathbf{Q}_s \mathbf{U}
   \left\{ a_{1,0},a_{2,0}, a_{3,0}, a_{4,0}\right\}^T
\end{equation}

As such, we can write the state vector as functions of the coefficients of the general solution, which yields the transmitted and reflected wave coefficients for a given known input incident wave. Mathematically speaking, equation (\ref{eq:Tn}) can be alternatively expressed as:
\begin{equation}
    \left\{  a_{1,2n}, a_{2,2n}, a_{3,2n}, a_{4,2n}\right\}^T
    =\mathbf{\Phi} 
 \left\{ a_{1,0},a_{2,0}, a_{3,0}, a_{4,0}\right\}^T
\end{equation}
where
\begin{equation}
    \mathbf{\Phi}
    =(\mathbf{E}_s \mathbf{Q}_s \mathbf{U} \mathbf{\Lambda}_n)^{-1} \mathbf{T}_n \mathbf{E}_s \mathbf{Q}_s \mathbf{U}
    =
    \begin{bmatrix}
        \mathbf{\Phi}_{11} & \mathbf{\Phi}_{12} \\
        \mathbf{\Phi}_{21} & \mathbf{\Phi}_{22} \\
    \end{bmatrix}
\end{equation}
such that $\mathbf{\Lambda}_n = \mathbf{\Lambda} (k_s,\ell_t)$ and $\mathbf{E}_s$ and $\mathbf{Q}_s$ are evaluated based on the substrate's mechanical properties. Note that $\ell_t = n\ell_p + (n-1)\ell_g$ is the total length of the metastructure segment, measured from $x =0$ that is fixed at the first metastructure interface (see figure~\ref{fig:each_patch_power}). Knowing that in the transmitted region, $a_{3,2n} = a_{4,2n} = 0$, and assuming an incident propagating wave with $a_{1,0} = a$ and $a_{2,0} = 0$, we can compute the complex wave amplitudes of transmitted and reflected waves as follows:
\begin{equation}
\begin{Bmatrix}
 a_{1,2n}  \\ a_{2,2n}
    \end{Bmatrix}
    = \left(\mathbf{\Phi}_{11}-\mathbf{\Phi}_{12} \mathbf{\Phi}_{22}^{-1} \mathbf{\Phi}_{21} \right)
\begin{Bmatrix}
    a \\ 0
    \end{Bmatrix}
\end{equation}
\begin{equation}
\begin{Bmatrix}
    a_{3,0}  \\ a_{4,0}
    \end{Bmatrix}
    = -\mathbf{\Phi}_{22}^{-1} \mathbf{\Phi}_{21}
\begin{Bmatrix}
    a \\ 0
    \end{Bmatrix}
\end{equation}

\section{Energy harvesting performance of electromechanical metastructure: a power flow approach}
\label{sec:performance}
\subsection{Power flow mathematical formulations}
In this section, we utilize the power flow approach to assess the energy harvesting performance. A schematic showing the power flow in the EMM is depicted in figure~\ref{fig:each_patch_power}. The power harvested by the piezoelectric units is of utmost importance in assessing the energy harvesting efficiency. By solving the voltage response for each bimorph piezoceramic layers via the TMM in section~\ref{Sec:TMM_Finite_MM}, the total time-averaged harvested power by the piezoelectric harvesters is computed:
\begin{equation}
\label{eqn:214}
P= \sum_{i=1}^{n} P_{i} = \frac{1}{2R}\sum_{i=1}^{n}  |{v}_{i}|^{2}
\end{equation}
where $P_{i}$ is the harvested power by the $i^{\text{th}}$ patch. The time-averaged mechanical power flow in the beam at any point can be calculated from: 
\begin{equation}
P_{\text{avg}}=\frac{1}{2}Re \left({\dot w}^{*}F-{\dot \phi}^{*} M\right)
\label{eq:avg_power}
\end{equation}
where ${\dot w}^{*}$ and ${\dot \phi}^{*}$ are the complex conjugate of velocity and angular velocity, respectively~\cite{Tol2016PiezoelectricEnhancement}. Based on equation~(\ref{eq:avg_power}), the incident power can be obtained as:

\begin{equation}
    P_\text{I} = YI_s k_s^3 \omega
\end{equation}
assuming a unit amplitude of incident wave, i.e. $a = 1$. Hence, the energy conversion efficiency of the EMM can be calculated as:
\begin{equation}
\label{eqn:215}
\eta = \frac{P}{P_{\text{I}}} = \frac{1}{P_{\text{I}}}\sum_{i=1}^n P_{i}
\end{equation}

A similar procedure can be followed to obtain the transmitted and reflected power ratios. It is important to note that the sum of the reflected, harvested and transmitted power ratios must be equal to unity at any given frequency.

\subsection {Piezoelectric harvester performance}
\label{sec:single_patch_opt}

When considering a single patch for harvesting traveling wave energy, three important points are emphasized in reference ~\cite{Tol2016PiezoelectricEnhancement}: (1) It is not possible to achieve a conversion efficiency higher than 50\% for a single bimorph harvester, as it is a reciprocal three-port system, (2) for a resistive-only circuit, the harvested power ratio reaches a local maximum when the patch length is approximately equal to an odd multiple of half-wavelength of the propagating wave, and (3) the inclusion of an inductor in the piezoelectric patch circuitry alleviates the aforementioned size dependence, making it capable of achieving peak harvesting capacity of 50\% near the resonance frequency of the electrical circuit.

\begin{figure}[h!]
\centering
\includegraphics[]{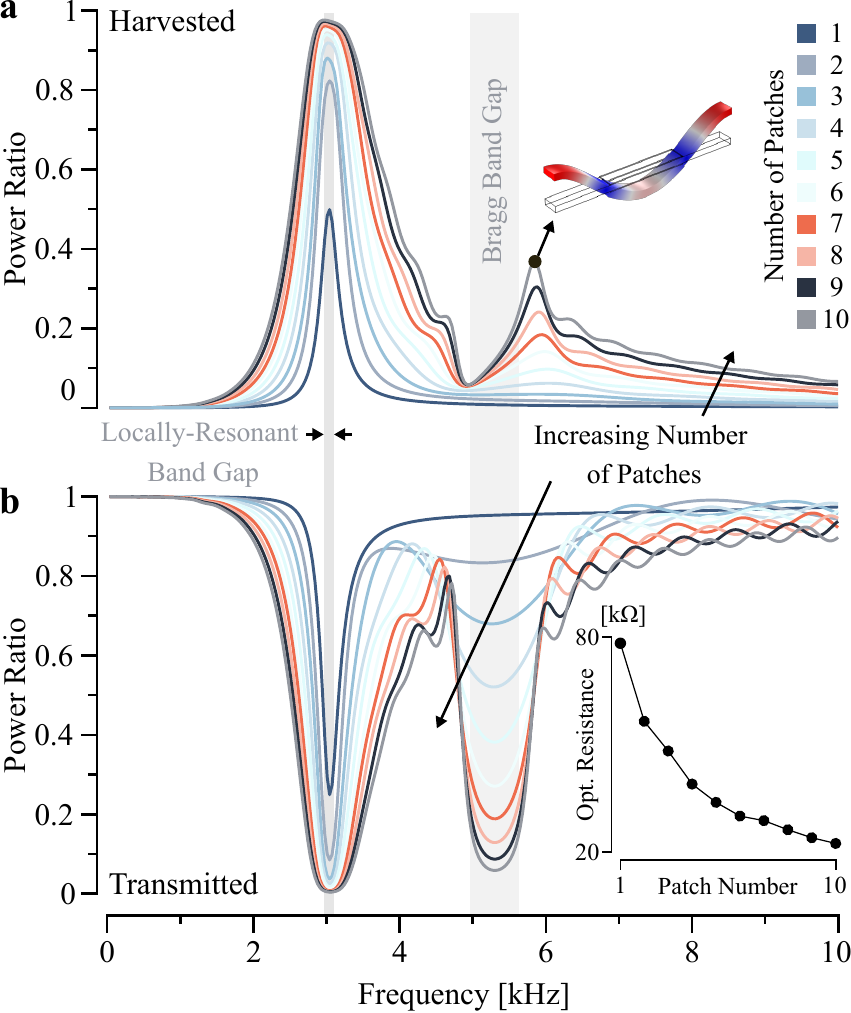}
\caption{Effect of the increase of the number of patches, $n$, on the (\textbf{a}) harvested and (\textbf{b}) transmitted power ratios. The inset in subfigure (\textbf{a}) demonstrates the mode shape of unit cell at the upper bounding frequency ($5.6$ kHz) of Bragg band gap. The inset in subfigure (\textbf{b}) depicts the optimal resistance found for each patch number based on the optimization procedure. Bragg and locally-resonant band gaps are highlighted in grey.}
\label{fig:increasing_number_patches}
\end{figure}

Next, we develop an optimization procedure of harvesting traveling wave energy in the EMM. For a targeted frequency $\omega_t=3.18$ kHz (corresponding wavelength $\lambda_t = 2\pi \sqrt[4]{\frac{YI_s}{m_s \omega_t^2}}$), a required inductance $L = (\omega_t^2 C_{p}^{eq})^{-1}= 200$ mH is chosen based on the measured equivalent capacitance of $C_{p}^{eq}=12.5$ nF. To obtain optimal harvested power ratios, we sweep a range of resistance values and record the resistance that yields the highest harvested power output near the targeted frequency, $\omega_t$. This procedure is performed for different number of patches. Figure~\ref{fig:increasing_number_patches}(\textbf{a}) shows that as the number of bimorphs increases, the peak value of harvested power ratio converges towards the maximum theoretical limit of $\eta = 1$. When $n$ becomes sufficiently large (which is approximately four patches here), the increase in the peak value becomes subtle with larger values of $n$, and its corresponding frequency slightly shifts to a frequency less than $\omega_t$. However, a larger number of patches is beneficial for a wider bandwidth of the energy harvester, which is clearly seen from the increase of area under the harvested power ratio curves of larger $n$ values. 

\begin{figure*}[]
\centering \includegraphics[]{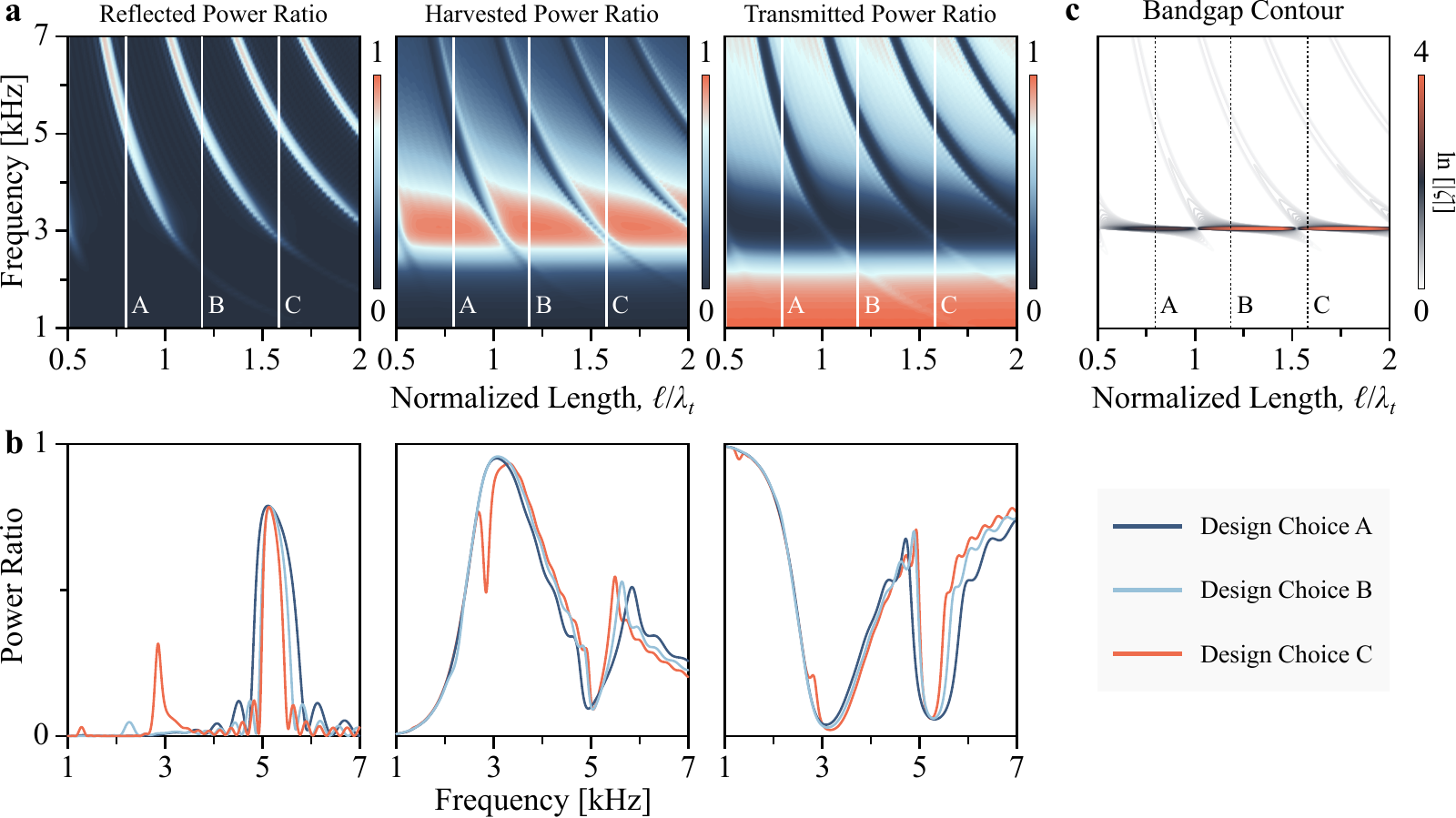}
\caption{(\textbf{a})  Reflected, harvested, and, transmitted power ratios under different excitation frequencies and unit cell sizes. Inductance and resistance of the electrical circuit and the patch length remain constant ($R=13$ $\text{k}\Omega$, $L=200$ mH, $l_p=25$  mm). (\textbf{b}) Choices of the EMMs (corresponding to the lines in subfigure (\textbf{a})), showing similar behavior. (\textbf{c}) Band gap contours for the swept range of normalized length, where $\ln[|\zeta|] \neq 0$ signifies a non-zero attenuation parameter and emergence of a band gap, and $\zeta$ stands for the eigenvalue of the transfer matrix as defined in section~\ref{sec:Transfer_Matrix_for_Electromechanical_Unit_Cell}.
}
\label{fig:Multiple_design_choices}
\end{figure*}

While a peak harvested power near $\omega_t$ is expected due to the electrical circuit resonance, a larger number of patches induces a distinguishable second peak near the upper bounding frequency of Bragg band gap. The second peak in harvested power plot starts within the Bragg band gap. It reaches a peak near its upper bounding frequency, indicating that some reflected wave energy due to Bragg band gap is recovered as harvested power in the EMM harvester, which further increases the operational bandwidth of the harvester beyond the conventional resonance frequency zone. We further explain the reason for the second peak by investigating the mode shape of the unit cell. Researchers have shown that bounding frequencies of the pass bands can be identified from the natural frequencies of a single unit cell with certain types of boundary conditions \cite{mead1975wave,tang2017broadband}. 
In our case, the mode shape corresponding to the upper limit of the Bragg band gap is shown in the inset in figure~\ref{fig:increasing_number_patches}(\textbf{a}), which is obtained from the unit cell analysis under periodic boundary condition using COMSOL multiphysics. It is clearly seen that the mode shape is
symmetric for the center of the unit cell so that the piezoelectric layers placed in the middle of the unit cell experience a large strain, resulting in a significant voltage output, thus enhancing energy harvesting performance. 

As shown in figure~\ref{fig:increasing_number_patches}(\textbf{b}), while the harvested power ratio steeply rises near the local-resonance frequency of $3.18$ kHz, the power transmission decreases as expected.~A similar trend is observed in the vicinity of $5.3$ kHz due to the Bragg band gap, and the transmitted power ratio decrease with larger number of piezoelctric bimorph, which is in agreement with previous observations in literature \cite{albabaa2017PC}. Hence, the multifunctional EMM can simultaneously achieve broadband elastic wave energy harvesting and vibration suppression capabilities.

\subsection{Multiple design choices}

\sloppy
The interest in this design analysis is to find an optimal resistance that gives significant peaks near local resonance and the upper limit of Bragg band gaps for different unit cell dimensions. Following the same optimization procedure, a resistance of 13~k$\Omega$ results in peak harvested power ratios of 0.95 and 0.51 near the local resonance and Bragg band gap, respectively, when a 10-bimorph metastructure with a gap length of $35.5$ mm is used. The corresponding frequencies of these peak values are found to be about $3.1$ kHz and $5.8$ kHz. Simultaneously, the broadband vibration mitigation performance based on 50$\%$ power attenuation is predicted as $1.8$ kHz and $1.1$ kHz in the vicinity of the band gaps in the multifunctional EMM system. Varying the unit cell's length ($\ell = \ell_p + \ell_g$) by sweeping the gap distance ($\ell_g$) while keeping the patch length ($\ell_p$) constant, such peak values can be guaranteed at different values of $\ell_g$ (and different unit cell's length as a result). At optimal resistance loading, the reflected, harvested, and transmitted powers are plotted in figure~\ref{fig:Multiple_design_choices}(\textbf{a}) for a range of frequencies and normalized unit cell lengths $\ell/\lambda_t$. It is interesting to see that the power ratio profile take a nearly periodic profile with the increase in the normalized length $\ell/\lambda_t$. This implies that the same performance of the energy harvester can be achieved at different geometrical scales while keeping the same targeted frequency, showing the design's versatility and potential in a wider range of applications. This is further demonstrated by the three examples at the three choices A, B, and, C, in figure~\ref{fig:Multiple_design_choices}(\textbf{b}), where all of them exhibit nearly an identical performance. Additionally, the reflected power maximize within Bragg band gap, the harvested power peaks at locally-resonant band gap, and the transmission is minimal whenever a band gap occurs, regardless of its type. The band gap contour (found based on the eigenvalues of the unit cell's transfer matrix) for the same swept range is provided in figure~\ref{fig:Multiple_design_choices}(\textbf{c}) for reference. 

A metastructure operating in the low-frequency range with large wavelength requires a compact design to avoid a bulky structure. Therefore, the design choice A is preferred in this study, given its relatively compact size and its significant harvested power ratios around the locally-resonant and Bragg band gaps. 

\subsection{Numerical verification}
\begin{figure}[h]
\centering
\includegraphics[]{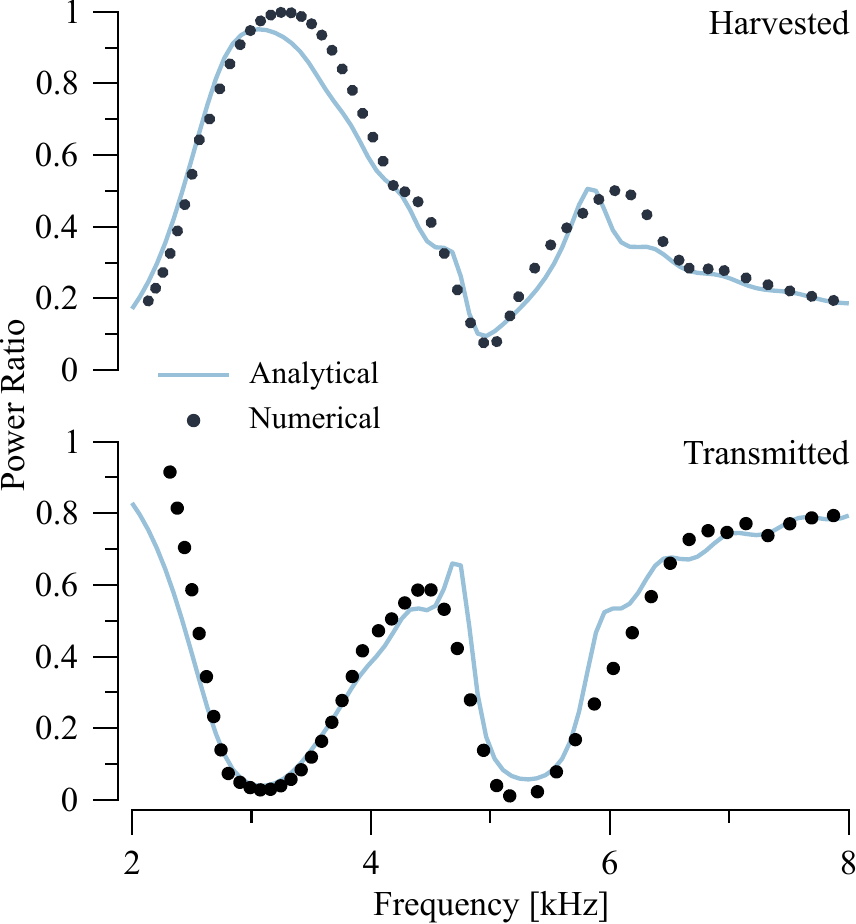}
\caption{Finite element simulation results for the harvested and transmitted power ratios, showing excellent agreement with analytical results.}
\label{fig:analytical_vs_numerical}
\end{figure}

To numerically verify the analytical framework, a corresponding finite element model is established, and time-domain simulations are conducted in COMSOL multiphysics. To ensure accurate finite element simulations, a high-quality mesh was used in the metastructure by setting the maximum mesh size to $\lambda/20$. Numerical simulations were performed by exciting the beam under four-cycle sine-burst with different center frequencies and recording the temporal response at each time step. The choice of the time step was based on a Courant-Friedrichs-Lewy number of 0.2 for optimal wave solution. Additionally, low-reflecting boundary condition was imposed at the edges to reduce the interference of boundary reflections.~Transmitted and harvested power ratios at different frequencies were then calculated by post-processing the temporal response using Fast Fourier Transform (FFT) technique. Figure~\ref{fig:analytical_vs_numerical} shows that the numerically obtained harvested and the transmitted power ratios are in a close agreement with the analytical predictions.

\sloppy
\section{Experimental validation}
\label{sec:experiment_validation}
\subsection{Experimental setup}
\begin{figure}[]
\centering
\includegraphics[]{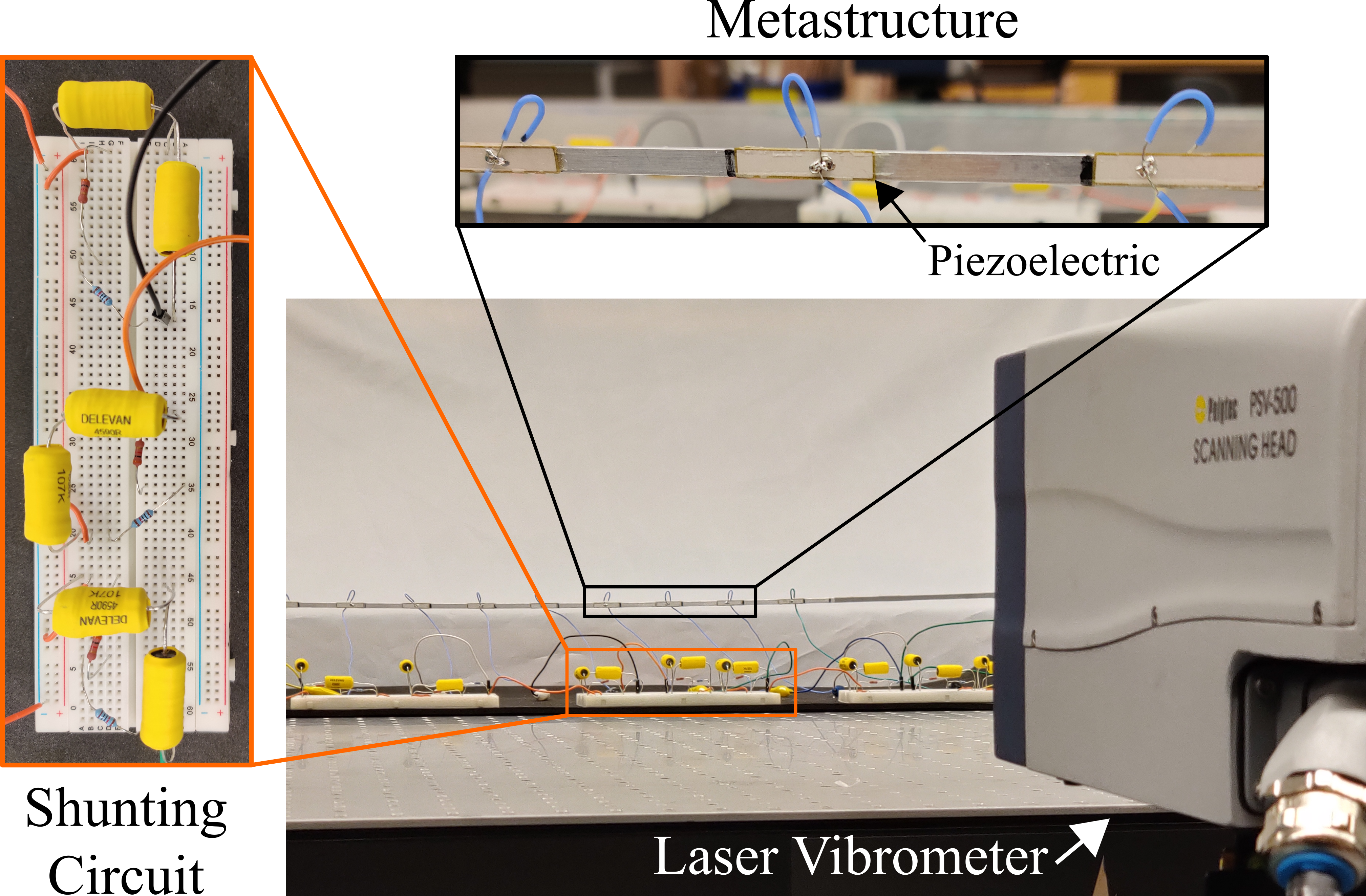}
\caption{Experimental setup, showing the scanning laser vibrometer head, metastructure with piezoelectric patches, as well as the shunting circuits.}
\label{fig:experimental_setup}
\end{figure}

The performance of the EMM is experimentally tested to validate the analytical and numerical results. Of primary focus here is to acquire the harvested and transmitted power ratios, which validates the energy harvesting performance and the wave attenuation phenomenon around locally-resonant and Bragg band gaps. Here, an EMM configuration identical to that of design choice A in figure~\ref{fig:Multiple_design_choices} is chosen for experimental testing. Ten piezoelectric patches (Steminc SM311, $25 \times 5 \times 0.3$ mm) are periodically placed at the mid-span of a 6 feet long beam, forming the metastructure section, and wired to their optimally configured circuitry (with inductance, $L=200$ mH, and resistance, $R=13$ $\text{k}\Omega$ in parallel connection). The length of the beam is sufficiently long to prevent the interference between the reflected wave from the boundary and the propagating wave in the metastructure region. The input incident signal was generated via a thickness-mode piezoelectric actuator of an identical type of the piezoelectric patches in EMM. The transverse velocity responses of the incident and transmitted waves were measured via a Polytec PSV-500 scanning laser Doppler vibrometer (LDV), while the voltage output of each piezoelectric bimorph cell was measured via a Tektronix TBS1064 oscilloscope. A picture of experimental setup showing its different components is presented in figure~\ref{fig:experimental_setup}. 
 
\begin{figure*}[h]
\centering
\includegraphics[]{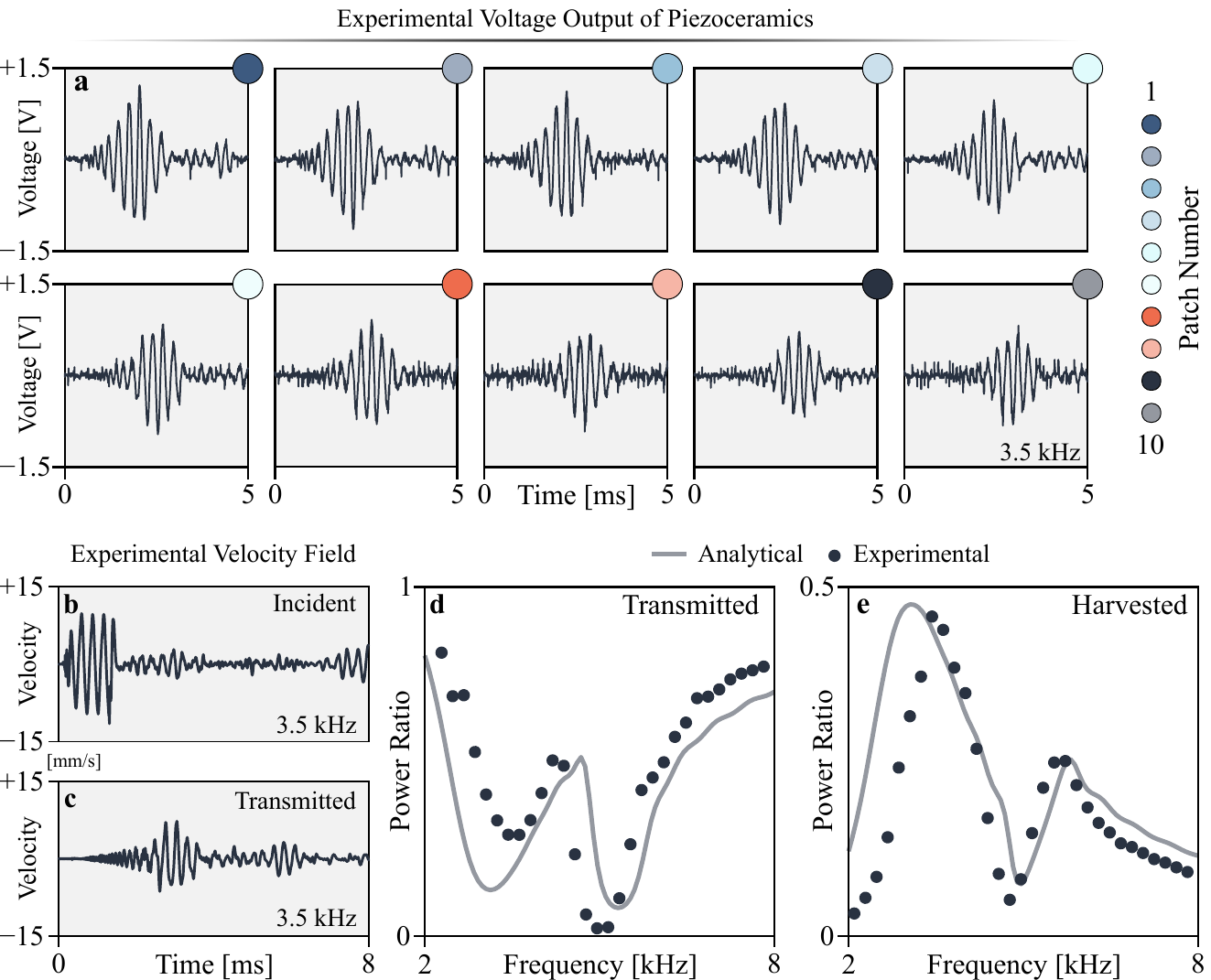}
\caption{(\textbf{a}) Experimentally measured voltage responses across the harvesters in the EMM under the excitation of four-cycle sine-burst at $3.5$ kHz central frequency. Velocity responses in (\textbf{b}) the incident and (\textbf{c}) the transmitted regions. (\textbf{d,e}) Comparison between experimental and analytical  transmitted and harvested power ratios showing a good agreement.}
\label{fig:experiemnt_signal_results}
\end{figure*}

\begin{figure*}[h]
\centering 
\includegraphics[]{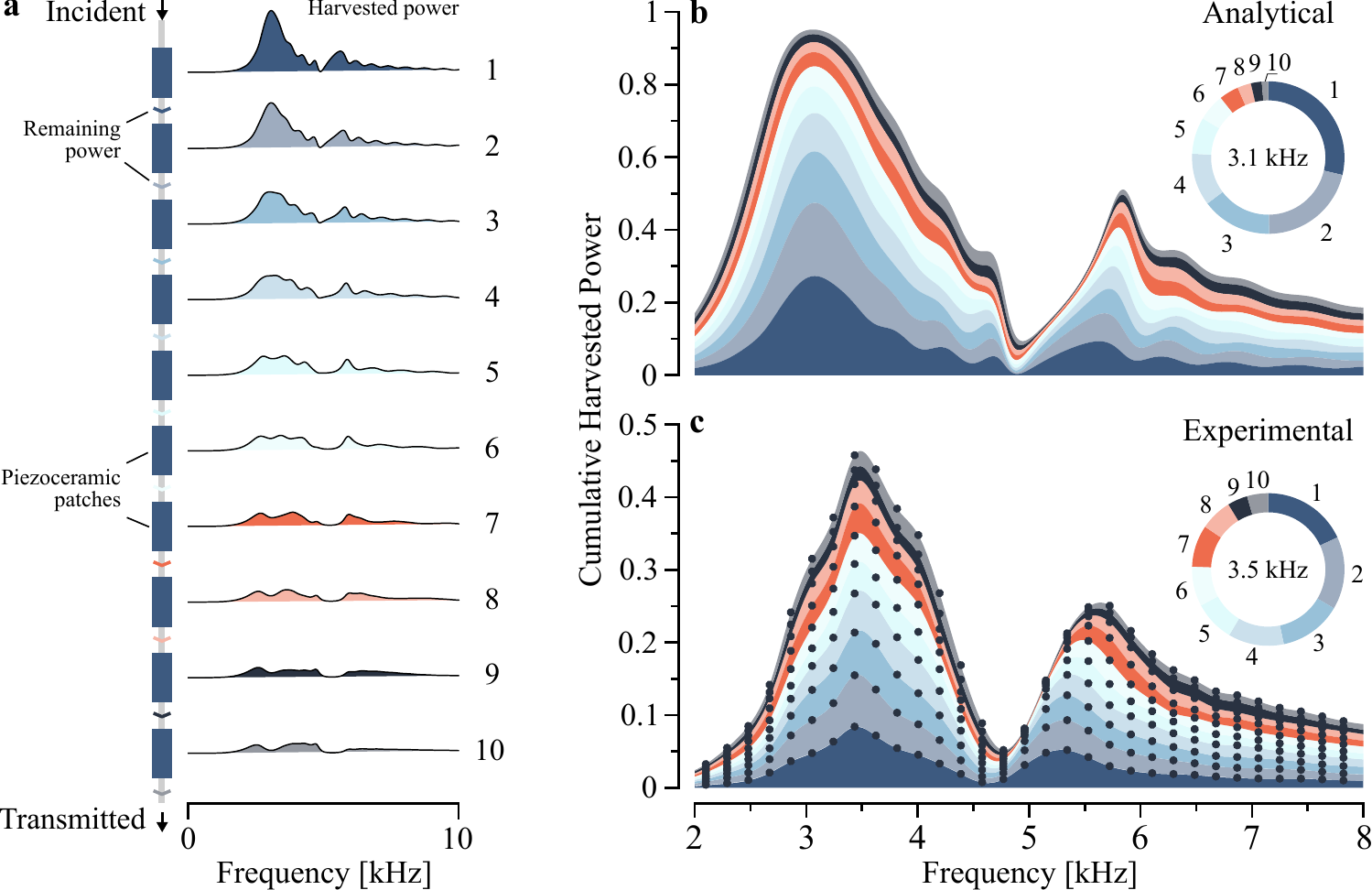}
\caption{ (\textbf{a}) Illustration of the analytical harvested power ratios by individual patches for a swept range of frequencies. Cumulative harvested power ratio from (\textbf{b}) analytical and (\textbf{c}) experimental investigations as the traveling wave passes through each unit cell (structural damping and dielectric loss are neglected for the analytical computations). The analytical (experimental) peak value appears near a frequency of $3.1$ $\text{kHz}$ ($3.5$ $\text{kHz}$) and the pie chart in the inset provides a visual presentation of the contribution of each patch from 1 to 10. The experimentally obtained harvested powers are depicted as circular dots.}
\label{fig:contribution_patches}
\end{figure*}

\subsection{Experimental results}

In the experiment, an incident flexural wave was generated with four-cycle sine-burst excitation at different central frequencies, ranging from $2$ to $8$ kHz using the thickness-mode piezoelectric actuator. Acquired velocity and voltage data were then post-processed using FFT technique to calculate the transmission and harvested power ratio (equivalently conversion efficiency). Considering the mechanical and electrical losses in the system, the structural damping factor ($\eta_s$), the dielectric loss tangent ($\tan(\delta)$), and the internal resistance of the inductors (measured as 140 $\Omega$) are introduced into the electromechanical model. The effect of the structural damping factor is introduced to the substrate's properties by assuming a complex Young's module of the material $\bar{Y}_s=Y_s(1+\textbf{i}\eta_s)$ and a complex flexural wavenumber $\bar{k}=k_s(1-\textbf{i}\eta_s/4)$, where $\eta_s$ is chosen as 1\%. In addition to the electrical losses from the internal resistance of the inductors, a complex permittivity $\bar{\varepsilon}_{33}^S=\varepsilon_{33}^S(1-\textbf{i} \tan(\delta))$ is used to account for the dielectric loss. The dielectric loss is estimated from the difference between the calculated and output conversion efficiency of the system, which is taken here as $15\%$.

\sloppy
Figure~\ref{fig:experiemnt_signal_results}(\textbf{a}-\textbf{c}) show experimentally measured voltage output, incident and transmitted velocities results at $3.5$ kHz, respectively. The locally-resonant and Bragg band gaps can be clearly identified in the transmitted power ratio shown in figure~\ref{fig:experiemnt_signal_results}(\textbf{d}). The experimental results are in good agreement with the analytical predictions when the mechanical and electrical losses are taken into account. The small discrepancies observed in the experimental transmitted power ratio are attributed to manufacturing imperfections in the mechanical (i.e., the bounding epoxy layer and wiring mass) and electrical components (i.e., extra resistance from wiring and epoxy). In particular, the attenuation observed in the locally-resonant band gap region is shifted from the analytical prediction (from $3.18$ kHz to $3.43$ kHz) and is believed to be caused by the mutual inductance effect from the adjacent inductors. Such an effect yields slightly reduced operational inductance, thus resulting in a higher local resonance frequency.

\sloppy
Also, based on the voltage responses obtained for the range of $2$ to $8$ kHz, the harvested power ratio is computed, where a good agreement with the analytical response is observed (see figure~\ref{fig:experiemnt_signal_results}(\textbf{e})).~A maximum harvested power ratio of 0.45 (at about $3.5$ kHz) and 0.25 (at about $5.6$ kHz) around locally-resonant and Bragg band gaps, respectively, are obtained. Such peak values are lower than the ideal values of 0.95 and 0.51, which are calculated when losses are neglected. Along with the transmission response, harvester efficiency successfully demonstrates that the proposed EMM system can simultaneously achieve a broadband energy harvesting and wave attenuation.

\subsection{Contribution of each patch}
\label{sec:patch_contribution}

\sloppy
As mathematically depicted in equation~(\ref{eqn:214}), the total harvested power from the metastructure system is a collective contribution of the array of patches. Thus, it is of interest to understand the individual contribution of each patch relative to the total harvested power ratio as presented in figure~\ref{fig:contribution_patches}(\textbf{a}). Figure~\ref{fig:contribution_patches}(\textbf{b}) shows a series of curves of the ideal cumulative harvested power results as it evolves from the first patch to the tenth patch (without considering the structural damping and dielectric loss in the analytical model). It is observed that the first two bimorphs play the most important role in the harvesting system and constitute about 50\% of the total harvested power at the peak frequency around $3.1$ kHz, as visually conveyed by the pie chart in the figure's inset. The third and fourth patch constitutes a little above 25\%, leaving the rest of the patches to collectively contribute to less than 25\%, as the majority of power has been already extracted. A similar trend is observed in the experimentally obtained bimorph's contribution to the final harvested power ratio, as shown in figure~\ref{fig:contribution_patches}(\textbf{c}). The peak harvested power ratio in the experiment occurs around $3.5$ kHz and the pie-chart inset reveals the contribution from each of the patches. It is seen that nearly 47\% contribution is obtained from the first three patches, 29\% for the next three and, lastly, 24\% for the last four patches. These results emphasize the importance of the first few bimorph cells in the overall energy harvesting performance, which is in a good qualitative agreement with the analytical results in figure~\ref{fig:contribution_patches}(\textbf{b}). This phenomenon can also be observed from figure~\ref{fig:experiemnt_signal_results}(\textbf{a}), where the voltage outputs across the bimorph cells closer to the actuator exhibit higher voltage amplitudes than the later ones.

\section{Concluding remarks}
\label{sec:concluding}
In this paper, a fully coupled electroelastic framework of a piezoelectric metastructure is introduced for harvesting and attenuation of travelling wave energy. Key concept of the study is the optimization of the mechanical power flow in the metastructure by exploiting the locally-resonant effects induced by the electrical resonance of the piezoelectric bimorphs connected to a resistive-inductive load and the Bragg-scattering effects resulting from the spatial periodicity of the unit cells. Broadband harvesting and wave attenuation in the metastructure are achieved through a comprehensive optimization procedure by selecting the optimal inductance and resistance values as well as the overall unit cell size normalized by the wavelength. Analytical and numerical results show that the conversion efficiency reaches up to $95\%$ at the local resonance frequency. Furthermore, a conversion efficiency of $51\%$ is obtained in the vicinity of the Bragg band gap frequencies showing a significant recovery and conversion of the reflected wave energy in the piezoelectric harvester units. Furthermore, wideband vibration mitigation performance based on 50$\%$ power attenuation is predicted as $1.8$ kHz and $1.1$ kHz in the vicinity of the band gaps. With the consideration of the losses in the electromechanical structure, the theoretical results are in good agreement with the experimental measurements. Hence, the proposed electromechanical metastructure offers a multifunctional design that can achieve simultaneous elastic wave energy harvesting and attenuation capabilities at a wide range of frequencies.

\section*{Acknowledgement}
This work was partially supported by the National Science Foundation under Grant No. CMMI-1933436.

\bibliographystyle{unsrt}

\bibliography{EMM}

\end{document}